\documentclass[floatfix,aps,prx,superscriptaddress,reprint,footinbib]{revtex4-1}

\usepackage[T1]{fontenc}
\usepackage[utf8]{inputenc}
\usepackage{lmodern}
\usepackage{amsmath}
\usepackage{amssymb}
\usepackage{siunitx}
\usepackage{physics}
\usepackage{xfrac}
\usepackage{bm}
\usepackage{graphicx}
\usepackage{xcolor}
\usepackage{hyperref}
\usepackage{ulem}

\graphicspath{{}, {../images/}, {../figures/} }

\begin{document}
\title{Time-frequency encoded single-photon sources and broadband quantum memories based on a tunable one-dimensional atom }


\author{Ilan Shlesinger}
\affiliation{Laboratoire Charles Fabry, Institut d'Optique, CNRS, Université Paris-Saclay, 91127 Palaiseau CEDEX, France}

\author{Pascale Senellart}
\email[E-mail: ]{pascale.senellart-mardon@c2n.upsaclay.fr}
\affiliation{Centre de Nanosciences et de Nanotechnologies, CNRS, Université Paris-Sud, Université Paris-Saclay, 91120 Palaiseau, France}

\author{Loïc Lanco}
\affiliation{Centre de Nanosciences et de Nanotechnologies, CNRS, Université Paris-Sud, Université Paris-Saclay, 91120 Palaiseau, France}
\affiliation{Université Paris Diderot, Paris 7, 75205 Paris CEDEX 13, France}

\author{Jean-Jacques Greffet}
\email[E-mail: ]{jean-jacques.greffet@institutoptique.fr}
\affiliation{Laboratoire Charles Fabry, Institut d'Optique, CNRS, Université Paris-Saclay, 91127 Palaiseau CEDEX, France}


\date{\today}

\begin{abstract}
A one-dimensional atom--- an atomic system coupled to a single optical mode--- is central for many applications in optical quantum technologies. Here we introduce an effective one-dimensional atom consisting of two interacting quantum emitters coupled to a cavity mode. The dipole-dipole interaction and cavity coupling gives rise to optical resonances of tunable bandwidth with a constant mode coupling. Such versatility, combined with a dynamical control of the system, opens the way to many applications. It can be used to generate single photon light pulses with continuous variable encoding in the time-frequency domain and light states that show sub-Planck features. It can also be exploited to develop a versatile quantum memory of tunable bandwidth, another key ingredient for quantum networks. Our scheme ensures that all above functionalities can be obtained at record high efficiencies. We discuss practical implementation in the most advanced platform for quantum light generation, namely the semiconductor quantum dot system where all the technological tools are in place to bring these new concepts to reality. 
\end{abstract}


\maketitle

\vspace {1 cm}

\section{\label{intro}Introduction}
A single atom coupled to a single mode of the electromagnetic field, also called a "one-dimensional atom"~\cite{1datom}, is a long sought system to generate and manipulate quantum light~\cite{atominterface}. Single atoms are known to interact with a single photon~\cite{Kimble1977,Grangier_1986}, a property that can be exploited to develop efficient single photon sources or to fabricate devices showing non-linearities at the single photon level. This requires however that the atom interacts mostly with a single mode of the electromagnetic field, so as to collect every emitted photon or to ensure the deterministic interaction of a single photon with the atom. 

Highly efficient atom-photon interfaces have been demonstrated with natural and artificial atoms using geometrical approaches~\cite{Lee2011}, or controlling the emitter spontaneous emission~\cite{Purcell1946} through inhibition---as demonstrated for emitters in very thin nanowires~\cite{diamond_nanowire, Claudon2010}--- or through acceleration of spontaneous emission--- as shown using various types of optical microcavities~\cite{Madsen2014,Gazzano2013}. Such systems have led to the demonstration of ultra-efficient sources of indistinguishable single photons~\cite{Somaschi2016,He2016,FiberedPillarSPS,Senellart2017} that allows scaling-up intermediate quantum computing tasks~\cite{Loredo2017,wang2017} and have shown non-linearities at the single photon level~\cite{Desantis2017,UnconventionalPhotonBlockade}, a critical step for efficient photon gates and quantum relays.

These one-dimensional atom structures however present a fixed spectral bandwidth, which defines once and for all the temporal profile of the emitted photons and the temporal window for the non-linear gates. This spectral bandwidth is determined by the modified spontaneous emission rate that is engineered through the Purcell effect (for cavity based structures) or dielectric screening (for nanowire based structures). In the case of cavity structures, one can continuously tune the atom bandwidth by controlling its detuning with the cavity mode, yet at the expense of reduced mode coupling. 

\begin{figure*}[t]
	\includegraphics[width=\textwidth]{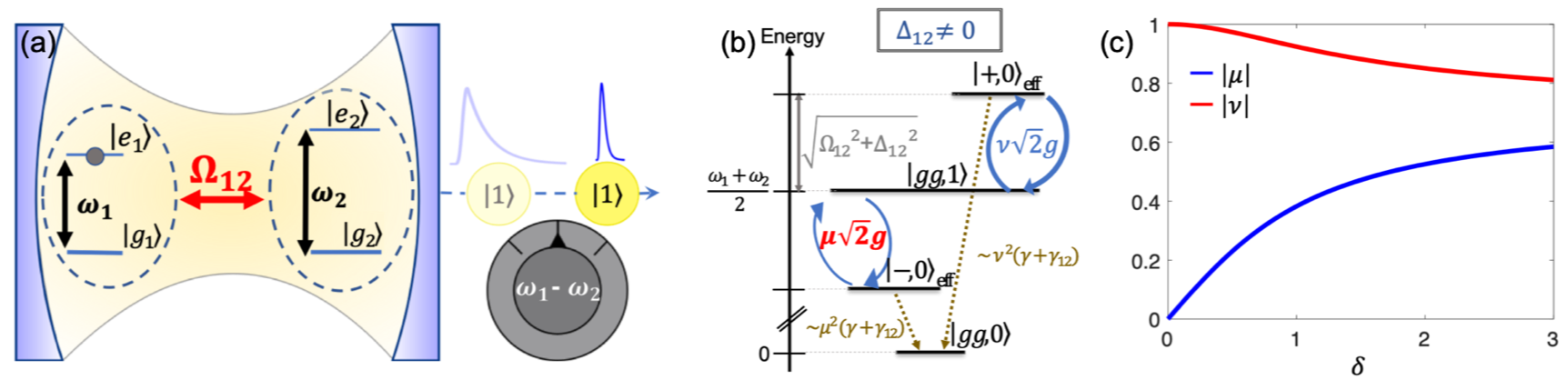}%
	\caption{\label{fig:system} (a) Two two-level systems (TLS) are weakly coupled to the same cavity mode. They are coupled to each other through dipole-dipole interaction with a rate $\Omega_{12}$ and their frequency mismatch $\Delta_{12}(t)=\omega_{1}-\omega_{2}$. 
	(b) Energy diagram of the eigenstates with at most one excitation. The first and second part of the kets correspond respectively to the atomic state and number of photons in the cavity. Double arrows represent coupling rates between the coupled-QD eigenstates and the cavity mode, single dotted arrows the decay rates into all other modes. Energy splittings are indicated in the energy scale. The cavity frequency is set here to the mean QD frequency $\omega_{0}=\frac{\omega_{1}+\omega_{2}}{2}$. (c) Modulus of the coefficients $\mu$ and $\nu$ corresponding to the subradiant and superradiant parts of the  eigenstates as a function of $\delta=\frac{\Delta_{12}}{\Omega_{12}}$. See text for details.}
\end{figure*}

In the present work, we propose an effective one-dimensional atom that offers bandwidth tunability with a mostly constant mode coupling to the cavity. We consider the system made of two atoms mutually coupled through dipolar interaction~\cite{Ficek1986,Sitek2009,deLeseleuc2017} and coupled to the same cavity mode. In the bad cavity limit, the eigenstates are superpositions of sub-radiant and super-radiant states with controllable weight~\cite{Shlesinger2018}. Remarkably, while the eigenstates present bandwidth tunability, their coupling to the cavity remains mostly constant. 
These unique features open many possibilities for optical quantum technologies where one can dynamically tune the system parameters during the emission or absorption of light while remaining in the one-dimensional atom regime. We first demonstrate the generation of single photon wavepackets with continuous variable encoding in the time frequency domain, light states that have recently attracted strong interest for quantum computing and quantum sensing~\cite{timefrequencyCVcomputing,subplanck}.  Single photon wavepackets in the form of cat and compass states are generated, both presenting sub-Planck features. We then propose a bandwidth tunable quantum memory, that stores short photon pulses and releases them with  tunable bandwidth. Such device shows great potential for quantum networks, be it for interfacing various quantum systems~\cite{Akopian2011,Kurizki2015,Meyer2015} or for synchronization~\cite{synchro1,synchro2,synchro3}. 

\noindent  The new concepts introduced here are applicable to all kinds of coupled natural or artificial atoms. They  are ready to be implemented in the semiconductor platform for quantum light generation  based on  quantum dots, where all basic ingredients have individually been demonstrated ~\cite{Somaschi2016,He2016,FiberedPillarSPS,Senellart2017,Widhalm2018}.

\section{A tunable one dimensional atom}\label{sec:system}


The probability for an atom to emit into a desired optical mode, also referred to as "mode coupling", scales as $\beta=\frac{\Gamma_{0}}{\Gamma_{0}+\gamma}
$ where $\Gamma_0$ ($\gamma$) is the spontaneous decay rate into the cavity mode (other modes). The most common way to obtain a one-dimensional atom ($\beta\approx 1$) is thus to control its spontaneous emission. High $\beta$ can be obtained either by reducing $\gamma$, i.e. inhibiting the spontaneous emission in the other modes, or by increasing $\Gamma_0$, i.e. accelerating the emission in the chosen mode. In both cases, changing the atom bandwidth $\Gamma_{0}+\gamma$ automatically modifies its mode coupling~\cite{Laucht2009,Faraon2010,Petruzzella2015,Kuklewicz2012}. We hereafter introduce an effective one dimensional atom where the system bandwidth and mode coupling are largely independent.

%

We consider the system consisting of two atoms represented by two level systems (TLS) strongly coupled through a dipole-dipole interaction and coupled to the same cavity mode~\cite{Shlesinger2018}. The TLS have ground states $\ket{g_{i}}$ and excited states $\ket{e_{i}}$ ($i=1,2$), with transition energies $E_{\ket{e_{i}}}-E_{\ket{g_{i}}}=\hbar\omega_{i}$ and a dipole-dipole coupling strength $\Omega_{12}$~\cite{Ficek1986}. When both TLS are degenerate, $\Delta_{12}=\frac{\omega_{1}-\omega_{2}}{2}=0$, the atomic eigenstates are the well-known collective states, i.e. the subradiant antisymmetric state $\ket{-}=\frac{\ket{e,g}-\ket{g,e}}{\sqrt{2}}$ and the superradiant symmetric state $\ket{+}=\frac{\ket{e,g}+\ket{g,e}}{\sqrt{2}}$. Both states are energy split by $2\Omega_{12}\simeq2\gamma\frac{3}{4\left(kd\right)^{3}}$
where $k=\frac{2\pi n}{\lambda}$ and $n$ is the index of the material for two parallel dipoles separated by a distance $d$ much smaller than the emission wavelength $\lambda$~\cite{Andrews2004,Lehmberg1970,Ficek2002}.

Introducing the symmetric $\sigma_{s}$ and antisymmetric $\sigma_{a}$ lowering operators, such that $\ket{\pm}=\sigma_{s,a}^{\dagger}\ket{gg}$, with $\ket{gg}=\ket{g_1}\otimes\ket{g_2}$, the coherent evolution of the system is given by the following Hamiltonian, written in the frame rotating at the mean frequency $\omega_{0}$ of the two TLS:
\begin{align}\label{eq:hamiltonian}
H(t)&=\Omega_{12}\left(\sigma_{s}^{\dagger}\sigma_{s}-\sigma_{a}^{\dagger}\sigma_{a}\right) + (\omega_{c}-\omega_{0})a^{\dagger}a
\nonumber\\
&+\Delta_{12}\left(\sigma_{s}^{\dagger}\sigma_{a}+
\sigma_{s}\sigma_{a}^{\dagger}\right)
+i\sqrt{2}g\left(a^{\dagger}\sigma_{s}-a\sigma_{s}^{\dagger}\right) ,
\end{align}
with $\omega_{0}=\frac{\omega_{1}+\omega_{2}}{2}$, $\omega_{c}$ the frequency of the cavity mode, $a$ the cavity field annihilation operator, and $g$ the coupling between one TLS and the cavity mode, taken equal for both TLSs.
A standard master equation for the density matrix of the system allows accounting for the coupling to the environment:
\begin{align}\label{eq:equation-maitresse}
\dot{\rho}=i\comm{\rho}{H}+\kappa\mathcal{L}(a)+\gamma_+\,\mathcal{L}\left(\sigma_{s}\right)+\gamma_-\,\mathcal{L}\left(\sigma_{a}\right),
\end{align}
where $\gamma_{\pm}=\gamma\pm\gamma_{12}$ and for a given operator $\hat{o}$, $\mathcal{L}(\hat{o})=\hat{o}\rho \hat{o}^{\dagger}-\frac{1}{2}\hat{o}^{\dagger} \hat{o}\rho-\frac{1}{2}\rho \hat{o}^{\dagger} \hat{o}$. $\kappa$ corresponds to the energy decay rate of the cavity and $\gamma$ is the spontaneous emission rate of each TLS into all the other modes. The collective states emission rates are either increased or reduced by $\gamma_{12}$ where $\gamma_{12}\simeq\gamma\Bigl(1-\frac{1}{5}(kd)^2\Bigr)$ is very close to $\gamma$ for $d<\lambda/2\pi n$~\cite{Lehmberg1970}. 

As shown by eq. \ref{eq:hamiltonian}, $\Delta_{12}$ acts as an effective coupling between the states $\ket{+}$ and $\ket{-}$. The system eigenstates $\ket{-}_\text{eff}$ and $\ket{+}_\text{eff}$ are then a superposition of both symmetric and antisymmetric states for which simple analytical expressions can be obtained when $\gamma\ll\Omega_{12},g$~\cite{Shlesinger2018}:
\begin{equation} \label{eq:detuned_coupled_modes}
\ket{-}_\text{eff}\simeq\mu \ket{+} + \nu \ket{-} ,\quad \ket{+}_\text{eff}\simeq\nu \ket{+} - \mu\ket{-} ,
\end{equation}
with
\begin{align} \label{eq:munu_analytic}
\mu&=\frac{\delta}{\sqrt{\delta^{2}+\left(1+\sqrt{1+\delta^{2}}\right)^{2}}} ,\nonumber \\
\nu&=\frac{1+\sqrt{1+\delta^{2}}}{\sqrt{\delta^{2}+\left(1+\sqrt{1+\delta^{2}}\right)^{2}}},
\end{align}
and $\delta=\frac{\Delta_{12}}{\Omega_{12}}$. The corresponding energy diagram is sketched in the Fig.~\ref{fig:system}(b) for $\omega_{0}=\omega_{c}$ and the modulus for $\mu$ and $\nu$ are shown in Fig.~\ref{fig:system}(c).




 The effective antisymmetric eigenstate $\ket{-}_\text{eff}$ couples to the cavity mode only through its symmetric component $\mu\ket{+}$ at a rate given by $g_{\ket{-}_\text{eff}}=\mu\sqrt{2}g$. This corresponds to its symmetric component, $\mu$, times the coupling rate of $\ket{+}$ with the cavity mode, $\sqrt{2}g$. This coupling can be tuned with $\delta$ from no coupling at all ($\mu\approx0$), i.e. a completely dark state, for $\Delta_{12}=0$, to the same coupling as a single QD when $\Delta_{12}\gg\Omega_{12}$ ($\mu\approx\frac{1}{\sqrt{2}}$ from Eq.~\ref{eq:munu_analytic}). Its emission rate into the cavity mode is then given by:
\begin{equation}\label{eq:Gamma-eff_mismatch}
\Gamma_{\ket{-}_\text{eff}}=\frac{4g_{\ket{-}_\text{eff}}^{2}}{\kappa}\frac{1}{1+\left(\frac{2\Delta_{c}}{\kappa}\right)^2}= \frac{8\mu^2 g^2}{\kappa}\frac{1}{1+\left(\frac{2\Delta_{c}}{\kappa}\right)^2}
\end{equation}
where the Fermi's golden rule gives rise to a dependence both on the coupling rate $g_{\ket{-}_\text{eff}}$ and on the final optical density of states, that depends on the detuning with the cavity mode $\Delta_{c}=\omega_{\ket{-}_\text{eff}}-\omega_{c}=\sqrt{\Delta_{12}^2+\Omega_{12}^2}-\Omega_{12}$~\cite{Andreani1999,Auffeves2010}.
 
The emission of the effective antisymmetric eigenstate into all modes other than the cavity mode takes place both through its symmetric and through its anti-symmetric component as shown by Eq.~\ref{eq:equation-maitresse}. Considering that when $d\ll\lambda$, we have $\gamma_{-}\simeq 0$ and $\gamma_{+}\simeq2\gamma$, this emission is mainly due to its symmetric component and is given as:
\begin{equation}\label{eq:gamma-eff}
\gamma_{\ket{-}_\text{eff}}\simeq\mu^{2}\gamma_{+}\simeq2\mu^{2}\gamma \, .
\end{equation}

Remarkably, the dipole-dipole interaction results in a modification of the emission rate in the cavity mode and in all the other modes both proportional to the symmetric part of the state $\mu^2$. As a result, the coupling to the mode of the effective antisymmetric state reads as:
\begin{equation}\label{eq:coop-eff}
\beta=\frac{\Gamma_{\ket{-}_\text{eff}}}{\Gamma_{\ket{-}_\text{eff}}+\gamma_{\ket{-}_\text{eff}}}= \frac{F_{p}}{F_{p}+1+\left(\frac{2\Delta_{c}}{\kappa}\right)^2}.
\end{equation}
where we have introduced the Purcell factor $F_{p}=\frac{4g^2}{\kappa\gamma}$. As a result, as long as $ \left(\frac{2\Delta_{c}}{\kappa}\right)^2 \ll F_{p}+1$, the $\ket{-}_\text{eff}$ state remains a one dimensional atom, while having a highly tunable bandwidth. 

\begin{figure}
	\includegraphics[width=.9\columnwidth]{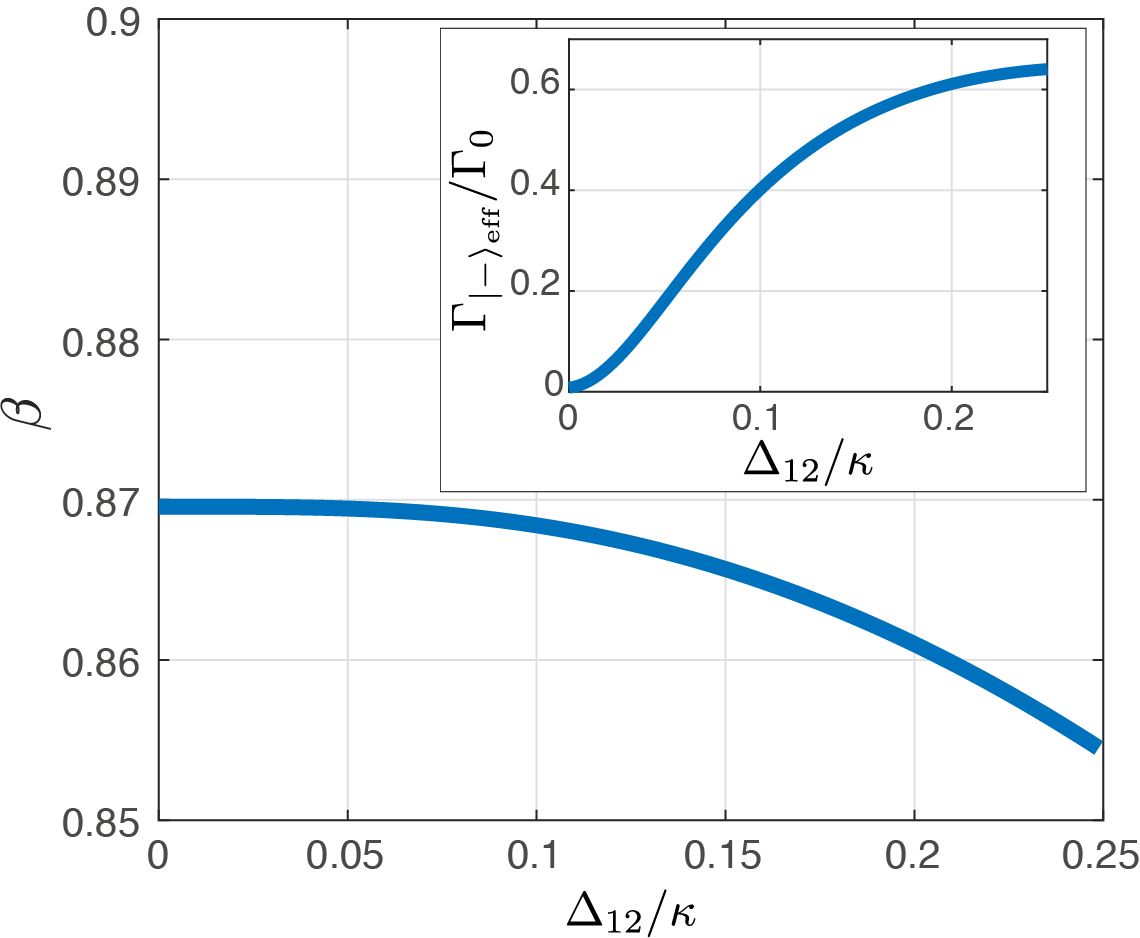}
	\caption{\label{fig:Gamma_beta} Coupling of the effective atom to the cavity mode  $\beta$ as a function of the detuning between the two  over the cavity linewidth. Inset: ratio of the spontaneous emission rate $\Gamma{\ket{-}_\text{eff}}$ of the subradiant state $\ket{-}_\text{eff}$ normalized to the spontaneous emission rate $\Gamma_0$ of a single TLS in the cavity. See the text for the parameters. Parameters corresponds to the case of two quantum dots in a micropillar cavity - see text for details.}
\end{figure}

To illustrate this effect, we now consider parameters corresponding to a technological platform that currently defines the state of the art in terms of emitter-based single photon sources~\cite{Senellart2017}, i.e. self-assembled InGaAs quantum dots (QDs) in micropillar cavities~\cite{Somaschi2016}. Yet the present concepts are also valid for other systems such as  color centers in diamond~\cite{Sipahigil2014} or QDs in photonic crystal cavities~\cite{Laucht2010} The values used here are $\left\{g,\kappa,\gamma\right\}=\left\{ 20, 400, 0.6 \right\}\,\mu\si{\electronvolt}$, corresponding to a system of high Purcell factor $F_{p}=6.6$ in the weak coupling or bad-cavity regime. We consider that the distance between the two QDs is $d=10\,\si{\nano\meter}$, a small distance that allows to neglect tunneling effects~\cite{Bayer2001} whilst providing a strong dipole-dipole coupling $\Omega_{12}=31\,\mathrm{\mu eV}$. With this condition, $\gamma_{12}=0.99\,\gamma$ where the spontaneous emission rate of a single QD in the leaky modes is $\gamma=0.6\,\mathrm{\mu eV}$. 

The spontaneous emission rate of the subradiant state $\Gamma_{\ket{-}_\text{eff}}$, normalized to the rate of a single QD in a cavity $\Gamma_{0}=\frac{4g^2}{\kappa}$, is plotted in the inset of Fig.~\ref{fig:Gamma_beta}. It is plotted as a function of the detuning between the two QDs, $\Delta_{12}$, normalized by the cavity linewidth $\kappa=400\,\mathrm{\mu eV}$. The cavity is set to the frequency of the subradiant state at zero detuning: $\omega_{c}=\omega_{\ket{-}}=\omega_{0}-\Omega_{12}$.
$\Gamma_{\ket{-}_\text{eff}}$ goes from $0$ for $\delta=0$ and increases with the detuning to reach $0.65\Gamma_{0}$ for $\Delta_{12}/\kappa=0.25$ evidencing a widely adjustable linewidth. In the same range of parameters, the main part of Fig.~\ref{fig:Gamma_beta} shows a mostly constant mode coupling, $\beta$, showing the persistence of the one-dimensional atom character over the whole range of detunings.

A straightforward application of such tunable one-dimensional atoms concerns an important challenge for quantum networks, namely, the bandwidth mismatch. Indeed,  quantum networks will most probably rely on interfacing various atomic systems, some of them providing the best quantum memories, such as atoms or ions, while others providing the most efficient single photon generation, such as QDs. A first experimental demonstration of QD-ion interfacing~\cite{Meyer2015} mitigated the large bandwidth mismatch between the QD and ion resonance by using a very low  coherent excitation of the QD. Such excitation, where the light only undergoes elastic scattering, provides single photons with the laser bandwidth but is limited to very low efficiency. The effective one-dimensional atom presented here allows a control of the single photon temporal profile at constant efficiency. In the case of QDs in micropillar cavities, the bandwidth ranges from around $10\,\mathrm{GHz}$ to $\mathrm{10\,MHz}$ for a mode coupling exceeding 85\%. As shown in the next sections, this system opens many more possibilities, such as the full engineering of the temporal profile of single photon wave packets or the possibility to develop a versatile quantum memory.

\section{\label{singlephoton} Time-frequency encoded single photon sources}

The system introduced here allows not only to control the spectral bandwidth of the atomic emission, but also its  phase and amplitude, and can eventually be used to obtain a full shaping of the single photon wavepackets~\cite{Keller2004,Thyrrestrup2013,Tufarelli2014} at constant high mode coupling. In the following, we discuss how to obtain bright single photon sources with the information encoded in the time-frequency domain~\cite{Brendel1999}.

\subsection{Tunable emission dynamics}

In the following, we consider that one can dynamically control the detuning $\Delta_{12}(t)$ between the two emitters. We model the spontaneous emission of our system restricted to the case of a single excitation. In the collective basis, we consider the states: $\left\{\ket{g,g,0}; \ket{-,0}; \ket{+,0}; \ket{g,g,1}\right\}$ where the first and second part of the ket correspond respectively to the two atomic states and number of photons in the cavity. In this basis the evolution of the mean values of single operators are described by a set of closed equations~\cite{Auffeves2008}:
\begin{align}\label{eq:evolution_op}
i\partialderivative{}{t}
\begin{pmatrix}
\ev{\sigma_{a}}\\
\ev{\sigma_{s}} \\
\ev{a} \\
\end{pmatrix}=
\begin{pmatrix}
-\Omega_{12}-i\frac{\gamma_{-}}{2} & \Delta_{12}(t) & 0 \\
\Delta_{12}(t) & \Omega_{12}-i\frac{\gamma_{+}}{2} & -i\sqrt{2}g\\
0 & i\sqrt{2}g & \tilde{\omega}_{c}-\omega_{0}\\
\end{pmatrix}
\begin{pmatrix}
\ev{\sigma_{a}}\\
\ev{\sigma_{s}} \\
\ev{a} \\
\end{pmatrix}.
\end{align}
where $\tilde{\omega}_{c}=\omega_{c}-i\frac{\kappa}{2}$. $\Delta_{12}$ is now a time dependent function. 

\begin{figure}
	\includegraphics[width=\columnwidth]{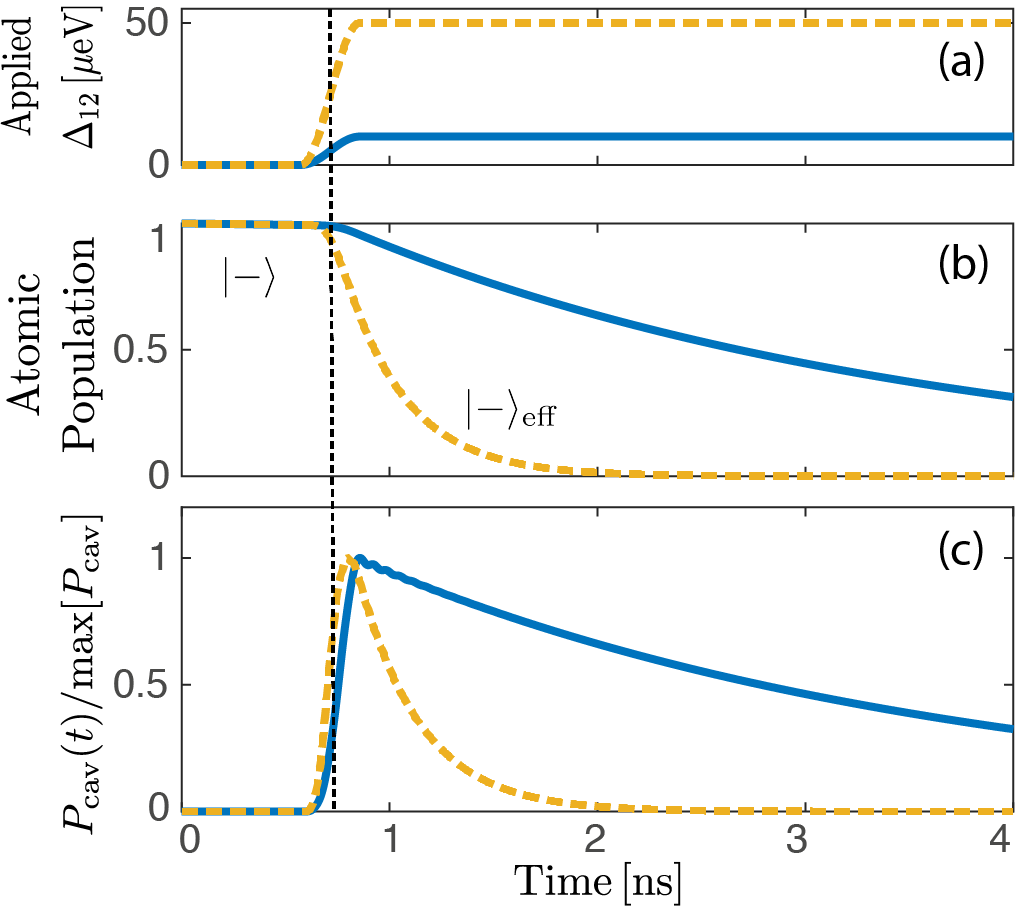}%
	\caption{\label{fig:tunable1} Controllable spontaneous emission. (a) Applied detuning $\Delta_{12}(t)$ as a function of time. (b) Total population of the two TLS. (c) Emitted power through the cavity (scaled). Solid blue corresponds to a maximal detuning of $\Delta_{12}=10\,\mu\si{\electronvolt}$ and dotted yellow to a maximal detuning of $\Delta_{12}=50\,\mu\si{\electronvolt}$. The vertical line across the three panels indicates the passage from the totally dark state $\ket{-}$ to the partially radiant effective antisymmetric state $\ket{-}_\text{eff}$ due to the applied detuning. The cavity is taken resonant to the final $\ket{-}_\text{eff}$ state in both cases: $\omega_{c}=\omega_{0}-\sqrt{\Delta_{12}^2+\Omega_{12}^2}$.}
\end{figure}

Fig.~\ref{fig:tunable1}~(a,b) illustrates the potential of $\Delta_{12}(t)$ dynamical tuning to control the emission time and emission dynamics. We consider that the system is in the subradiant state $\ket{-}$ at $t=0$ and that $\Delta_{12}=0$. 
The population of this subradiant state is uncoupled from the cavity and decays only slowly into the leaky modes at a rate $\gamma-\gamma_{12}$. Considering the previous parameters for the QD systems, it corresponds to a decay time of $1/\gamma_{-}=93\,\si{\nano\second}$, much slower than the natural decay time of around $1\,\si{\nano\second}$ for a QD in bulk. This slow decay is not visible on the time scale of Fig.~\ref{fig:tunable1}. 
To turn on the spontaneous emission, $\Delta_{12}(t)$ is set to a finite value to induce an effective coupling of the subradiant state with the symmetric, superradiant, state while keeping the phase relations between the two emitters.
This corresponds to plugging in the interaction with the cavity and hence Purcell enhancing the decay of the atomic state into the cavity mode.
The process is performed adiabatically, i.e. $\dot{\Delta}_{12}\ll\Omega_{12}^{2}$, so that the subradiant state $\ket{-}$ is converted only into the effective antisymmetric state $\ket{-}_\text{eff}$. Otherwise, if the applied detuning is too fast, the excitation will also be transferred to the symmetric state $\ket{+}_\text{eff}$, and will be emitted rapidly without the possibility to control the emission. Most importantly, such adiabatic coherent process ensures the indistinguishability of the emitted light wavepackets. 

The emission rate through the cavity mode depends on the symmetric portion of the $\ket{-}_\text{eff}$ state and hence on the final value of the detuning to which the TLSs are brought. The more detuned the TLSs are at the end, the more radiant  $\ket{-}_\text{eff}$ becomes and the faster is the decay. In Fig.~\ref{fig:tunable1},  a non-zero detuning is applied around $t=0.7\,\si{\nano\second}$ with a rise time equal to $280\, \mathrm{ps}$. Two examples are shown: the solid blue case with final detuning $\Delta_{12}=10\,\mu\si{\electronvolt}$ decays slower than the dotted yellow case with final detuning $\Delta_{12}=50\,\mu\si{\electronvolt}$.

The present approach requires a control of $\Omega_{12}$ and a dynamical tuning of $\Delta_{12}$. The dipole-dipole interaction strength depends on the distance between the TLSs, which can be thoroughly controlled during the growth process for QD structures. The detuning between the QD spectral resonances can be modified thanks to the confined differential stark effect by applying an external electric field~\cite{Unold2004,Kim2009,Sonnenberg2014}. This can be done with electrical pulses as fast as $100\,\si{\pico\second}$ and rising edges below $50 \,\si{\pico\second}$ as experimentally shown~\cite{Widhalm2018}.

\subsection{Generation of cat-like state and compass state single photon wavepackets}

We extend the control of the emission process to achieve complex pulse shaping of the single photon's amplitude. One can indeed obtain complex time structures for the emitted single photon wavepackets by turning on or off the coupling with the cavity mode through $\Delta_{12}(t)$ during the emission process. 

\begin{figure}
	\includegraphics[width=.99\columnwidth]{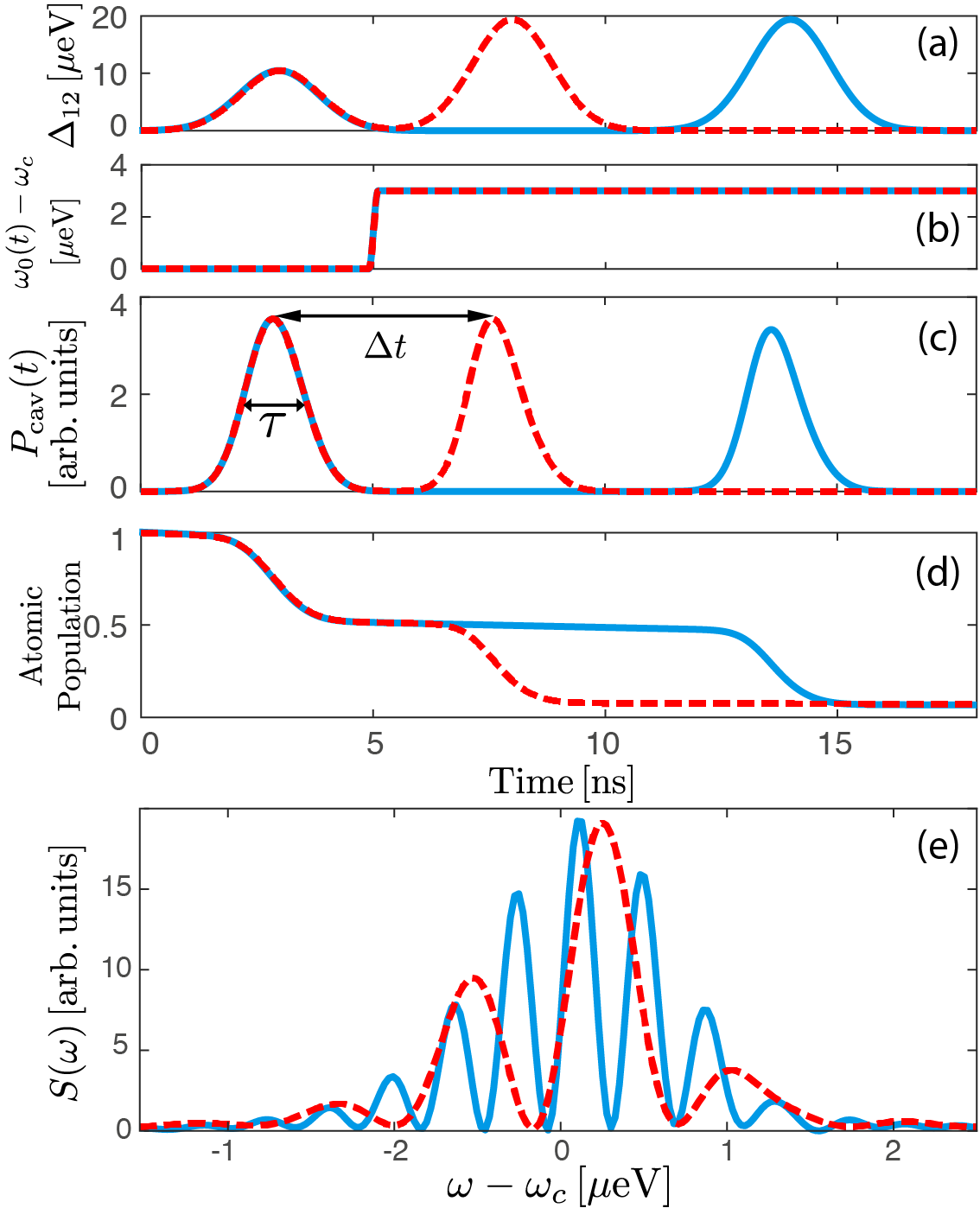}
	\includegraphics[width=.99\columnwidth]{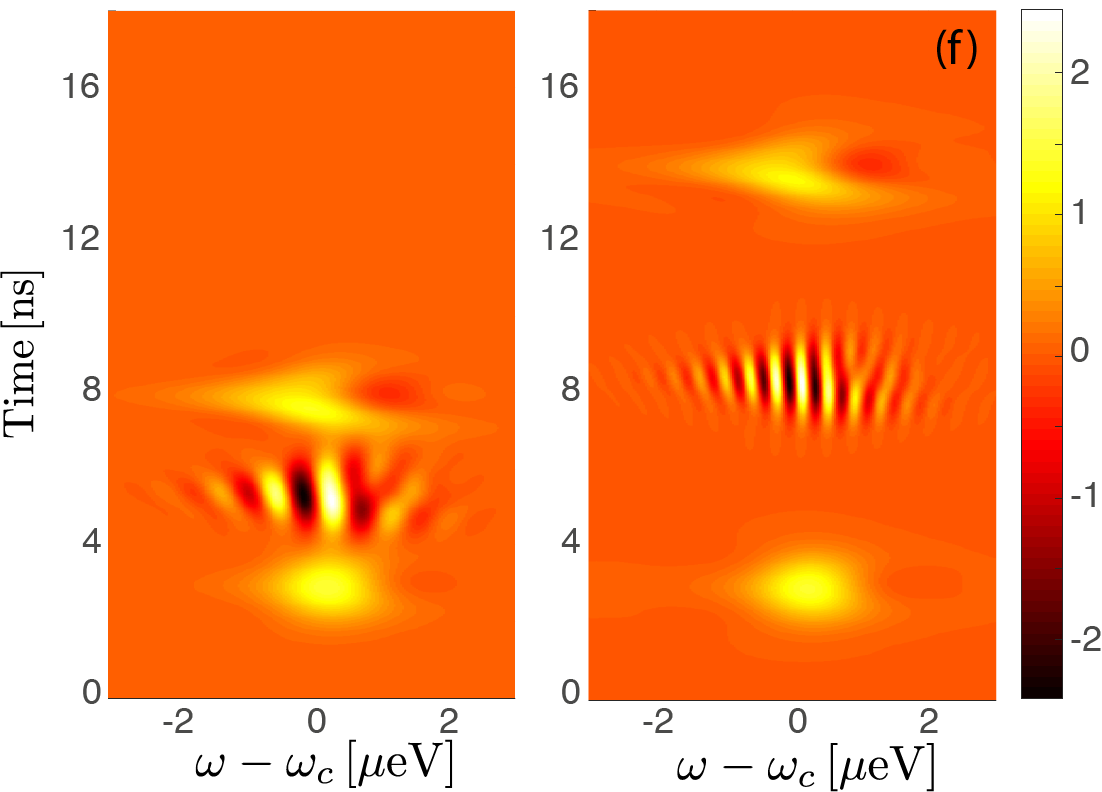}%
	\caption{\label{fig:peignes} Time bin encoding of a single photon. (a) Applied detuning sequence $\Delta_{12}(t)$ (b) and overall frequency shift $\omega_{0}(t)$. (c) Corresponding emitted power by the cavity. Two pulse sequences are considered with a waiting time between each pulse of $5\,\si{\nano\second}$ (dashed red) and $11\,\si{\nano\second}$ (dashed red). (d) Population of the collective state $\ket{-}_{\mathrm{eff}}$.   (e) Density power spectrum of the power emitted by the cavity. (f) Corresponding Wigner-Ville function of the emitted electric field: left waiting time of $5\,\si{\nano\second}$ and right waiting time of $11\,\si{\nano\second}$. }
\end{figure}

We first apply two $\Delta_{12}(t)$ pulses separated in time as shown in Fig.~\ref{fig:peignes}(a).
The mean value of the emitted power then shows two distinct peaks as shown in Fig.~\ref{fig:peignes}(c), corresponding to a time bin encoding of the information. The time interval between the peaks $\Delta t$ can be chosen at will and the duration of each peak $\tau$ is controlled by the maximum value of $\Delta_{12}$ which is applied. Two different examples are depicted with a waiting time $\Delta t=5\,\si{\nano\second}$ (in dashed red) and $\Delta t=11\,\si{\nano\second}$ (in solid blue), again considering experimental parameters for the QD system.
Note that to obtain the same emitted power for the two peaks, the maximum applied detuning is higher for the second peak. We also assume  that one can not only control the emitters detuning $\Delta_{12} (t)$ but also the mean frequency of the two emitters $\omega_{0}(t)$ as shown in Fig.\ref{fig:peignes}(b).


The corresponding energy spectral density is plotted in Fig.~\ref{fig:peignes}(e). As expected by Fourier transform, the two peaks in time-domain result in a frequency comb. The overall width is proportional to the inverse of the duration of one peak $1/\tau$, which is equal for the two different cases here.
The interval between the fast frequency oscillations is proportional to the inverse of the waiting time $1/\Delta t$. For the case where $\Delta t=11\,\si{\nano\second}$ one obtains fast oscillations and narrow peaks.

A more detailed insight into the generated single photon wavepacket can be obtained by calculating the associated Wigner-Ville function~\cite{Ville1948}, also called the chronocylic Wigner function~\cite{Paye1992}.%
\begin{figure}
	\includegraphics[width=\columnwidth]{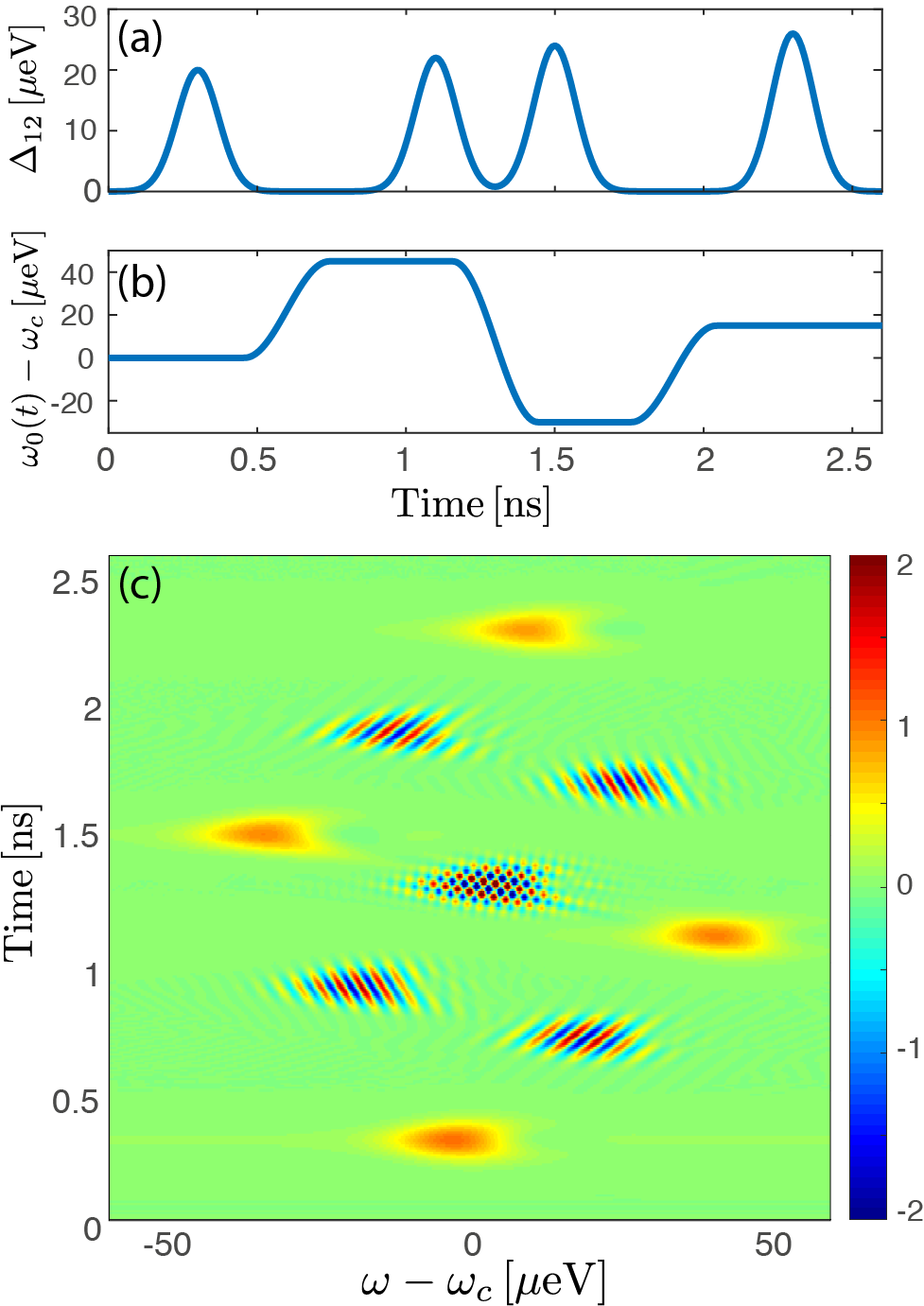}%
	\caption{\label{fig:compass} Generation of a compass state. The emission of such a state is obtained by applying a four-pulse detuning $\Delta_{12}(t)$ (a), and at the same time an overall frequency shift of the mean atomic frequency $\omega_{0}(t)$ (b). The cavity frequency is fixed at $\omega_{c}=\omega_0(0)-\Omega_{12}$. (c) Wigner-Ville function of the emitted electric field for a one-photon  state. }
\end{figure}
Its expression is given for a detection between $\left[0,T\right]$ by:
\begin{align}\label{eq:wigner}
W(t,\omega)=\frac{1}{2\pi}\int_{-\sfrac{T}{2}+\abs{t-\sfrac{T}{2}}}^{\sfrac{T}{2}-\abs{t-\sfrac{T}{2}}} d\tau \ev{a^{\dagger}(t+\frac{\tau}{2})a(t-\frac{\tau}{2})} e^{-i\omega\tau} .
\end{align}
It corresponds to the Fourier transform of the electric field correlations as a function of time and is displayed in Fig.~\ref{fig:peignes}(f) for the two different waiting times $\Delta t$. One recognizes a photonic cat-like state.
The oscillations visible at the mean time between the two emission peaks result from the interference between the probability to emit a photon at a time $t-\tau/2$ with the probability to emit a photon at $t+\tau/2$. They present time-frequency sub-Planck features, subtle features of quantum states that have recently been proposed for advanced quantum sensing applications~\cite{subplanck}. 

This first example illustrates the versatility of the tunable one-dimensional atom for the generation of light in two time bins. It can naturally be extended to more complex states of light since the effective atomic system presents a tunable lifetime and the phase of the state is controlled at the atomic level. This is illustrated in Fig.~\ref{fig:compass} where a set of four $\Delta_{12}(t)$ Gaussian pulses, separated in time are applied as plotted in Fig.~\ref{fig:compass}(a). The pulses are $170\,\mathrm{ps}$ wide and are applied at $t=\{300,\,1100, 1500,\,2300\}\,\mathrm{ps}$. 
The amplitude of each pulse is increased each time so as to compensate the smaller remaining emitter population.
In parallel, the mean frequency of the emitters, $\omega_{0}(t)=\frac{\omega_{1}+\omega_{2}}{2}$ is shifted to four different values as shown in Fig.~\ref{fig:compass}(b).
As seen in Fig~\ref{fig:compass}(c), a compass state, corresponding to a four-legs time-frequency superposition, is generated. It shows an even richer structure in the time-frequency domain, similar to the higher class of state that was recently introduced for sensing application in optomechanics \cite{andrewhypercube}.

Such structure is also ideal for continuous variable time encoding of the quantum information, a new field of quantum computing that exploits the continuous degrees of freedom of time encoding to enlarge the Hilbert space of computation \cite{timefrequencyCVcomputing}. 
Until now, the generation of continuous variable at the single photon level was limited to ultra-low source efficiency, whether it is based on parametric down-conversion \cite{spdcCVencoding1,spdcCVencoding2,spdcCVencoding3,spdcCVencoding4} or on the weak coherent excitation of an atomic system~\cite{Pursley2018,Matthiesen2013}. The approach introduced here, based on the dynamical control of a tunable one-dimensional atom, allows generating arbitrary time-encoded photons with near unity efficiency.



\section{A broadband and versatile quantum memory}

Another application of the present architecture is single photon storage and bandwidth conversion.
Many systems are explored to develop efficient quantum memories for quantum networks, most of them based on atomic vapors or ion-doped crystals~\cite{Bussieres2013}. They present very narrow bandwidths that prevent the efficient storage of short photon pulses, the latter being naturally desirable for high rate quantum networks. To date, the development of broadband single photon memories has been scarcely addressed. Such memory has recently been demonstrated using Raman assisted transition in warm atomic vapors: the proposed protocol offers interesting possibilities for releasing photons of improved quantum purity~\cite{IanBuffer}. However, so far the storage times remains in the few nanosecond timescale~\cite{Kaczmarek2018}. Here we show that, by dynamically tuning the emission lifetime of the subradiant state, one can achieve the storage of 100\,ps single photon pulses for over 100\,ns using an implementation of these concepts in the QD system. Naturally, building on the results of the previous section, the quantum information could also be released with a highly tunable temporal profile, independently of the incoming single-photon profile.



The storage mechanism relies on sending a single photon wavepacket on the tunable one-dimensional atom. The process takes place in three steps as shown in Fig.~\ref{fig:pump}(a). First the system parameters are tuned to a non-zero value for the detuning $\Delta_{12}$ so that they are effectively coupled to the cavity mode, allowing to absorb the incoming excitation, and excite the effective subradiant state $\ket{-}_{\mathrm{eff}}$. More specifically, $\Delta_{12}$ is chosen to match $\Gamma_{\ket{-}_\mathrm{eff}}$ with the bandwidth of the light pulse that one wants to store. The photon is then stored by tuning back the two QDs into resonance to adiabatically go from the effective subradiant state $\ket{-}_{\mathrm{eff}}$ to the dark state $\ket{-}$. The photon absorption process is then the time-reversed of the emission process starting with the system in the ground state $\ket{g,g,0}$ at $t=0$ and dynamically tuning the detuning $\Delta_{12}(t)$ so as to progressively bring the system into the fully antisymmetric, dark state $\ket{-}$ where it presents the minimal coupling to light. 
Finally, release  of the single photon is obtained by applying a detuning $\Delta_{12}$ between the two QDs.

While the incoming pulse is a single photon Fock state, it is easier to calculate this storage process using a coherent state $\ket{\alpha}=\sum_{k}\frac{\alpha^k}{k!}\ket{k}$ with a very low mean photon number $\ev{n}=\abs{\alpha}^2=0.01\ll1$. Only the one photon component excites the atom so that the system response reflects the response to single photon Fock state when normalizing the atomic population by the probability of having one photon~\cite{Kim2013,Giannelli2018}. The system excitation is done via the cavity mode through the Hamiltonian of Eq.~\ref{eq:hamiltonian}: $H_{p}(t)=-i\mathcal{E}_{p}(t)(a^{\dagger}-a)$, where $\mathcal{E}_{p}$ is the field amplitude coupled to the cavity mode.

Fig.~\ref{fig:pump} shows the evolution of the atomic population corresponding to a single photon Fock state sent on a QD system with the same parameters as in the previous sections. The single photon wavepacket to store presents a Gaussian temporal profile with a FWHM of $\tau_{\mathrm{pulse}}=550\,\mathrm{ps}$. The system parameters are set so that the incoming photon is resonant with the $\ket{-}_\text{eff}$ state:
\begin{align}
	\omega_{\ket{-}_\text{eff}}=\omega_{0}-\sqrt{\Delta_{12}^{2}+\Omega_{12}^{2}} .
\end{align}
\noindent The cavity frequency is tuned to $\omega_{c}=\omega_{0}-\Omega_{12}$ and $\Delta_{12}(t=0)$ is set to $40\,\mu\si{\electronvolt}$, in order to achieve the highest absorption probability. This condition is obtained when the $\ket{-}_\text{eff}$ state linewidth is equal to $\frac{2}{3}$ of the incident single-photon pulse width~\cite{Wang2011,Stobinska2009} (here it corresponds to $\mathrm{FWHM}=367\,\mathrm{ps}$). 
The QDs are then adiabatically brought into resonance: $\Delta_{12}(t)\rightarrow0$ during the storage process in order to bring the $\ket{-}_\text{eff}$ state into the darkest state $\ket{-}$. The photon will thus be stored for a time as long as $1/\gamma_{-}=93~\si{\nano\second}$ and recovered by setting $\Delta_{12}(t)$ to a non-zero value. 

Considering the present experimental parameters, we obtain a 68\% storage efficiency, given by the maximum flip inversion of the atomic system with a single photon pulse excitation. To stay within 5\% of such maximum efficiency, we find that the error in the start time of the detuning gate $\Delta_{12}(t)$ has to be smaller than 100\,ps which corresponds to a fifth of the single photon pulse. If the QDs are brought to resonance too early, the excitation stops before maximum population is achieved, and if the QDs are brought to resonance too late, spontaneous emission takes place before reaching the dark state. This value already represents a very high storage probability considering all current memory architectures where typical values are in the tenths of percent range~{\cite{Heshami2016}. 
The coupled quantum dot interface introduced here can thus be used to store short single photon pulses of 100\,ps for around 100\,ns, as shown in figure~\ref{fig:pump}(a), surpassing by almost two orders of magnitude the current state of the art~\cite{Poem2015,Kaczmarek2018}.
\begin{figure}
	\includegraphics[width=1\columnwidth]{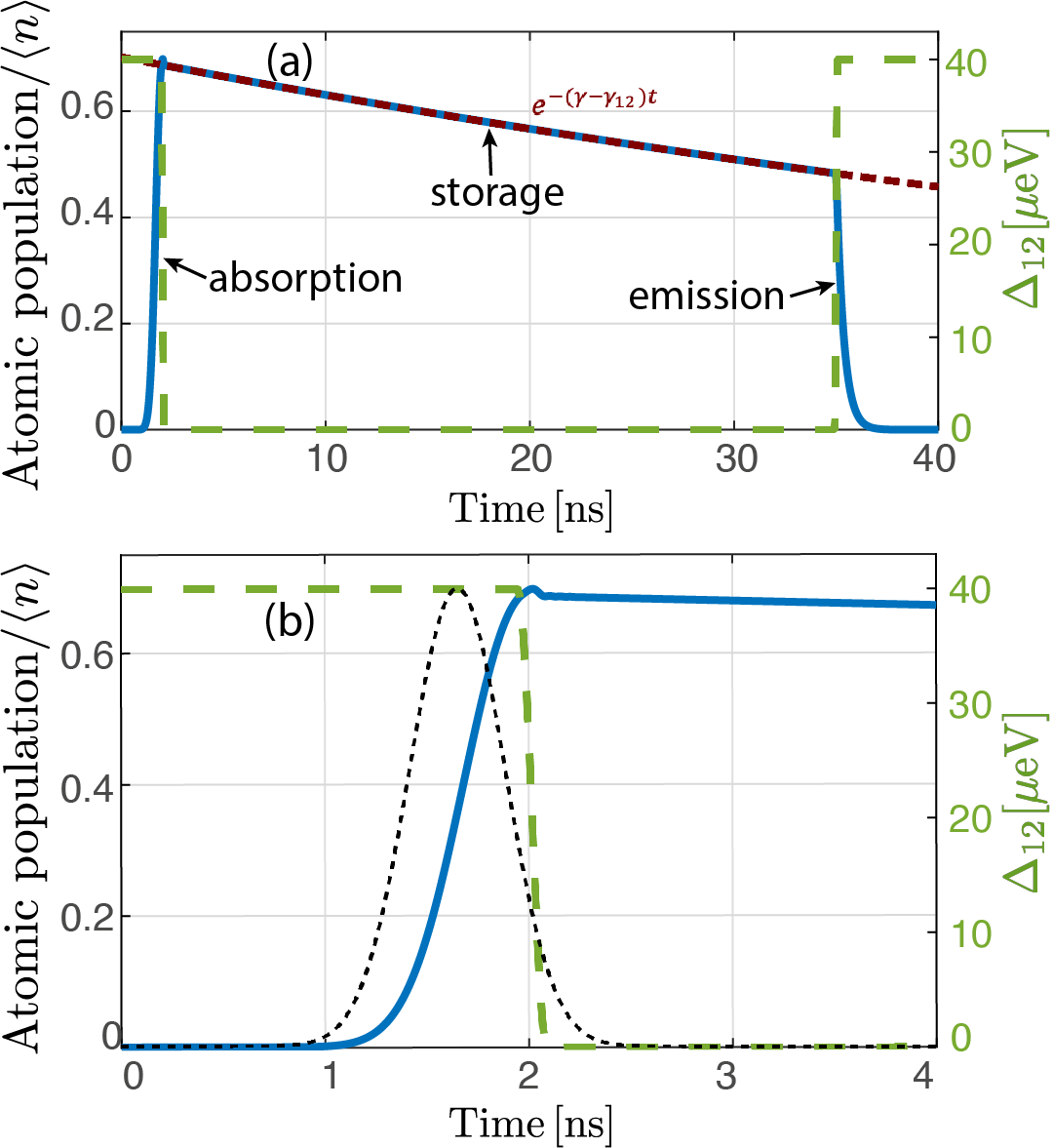}%
	\caption{\label{fig:pump} Broadband memory operation.	(a) Atomic population  normalized by the average number of photons in the pulse $\ev{n}$ (solid blue) and applied detuning ($\Delta_{12}\neq0$ in dashed green). The storage process takes place in three steps: first the two QDs are detuned so as to absorb the incoming photon , the excitation is then stored in the subradiant state by bringing the two QDs into resonance, and finally the photon is released by detuning again the two QDs at t=35~ns. The pump is resonant with the $\ket{-}_\text{eff}$ state.During storage, the excitation leakage takes place at a rate $\gamma_{-}=\gamma-\gamma_{12}$ (as shown in dotted brown).
	(b) Close-up of figure (a) at short time delays. The Gaussian profile of the incoming wavepacket is shown by the black dotted line. }
\end{figure}

We note that tunable single photon memories have recently attracted attention for their ability to re-emit the photon with a different frequency and bandwidth than the initial ones~\cite{Heshami2016}.
Indeed, frequency and bandwidth conversion is a valuable tool for building quantum networks.
A bandwidth modulation of between $0.5$ and $1.9$ of the incoming bandwidth was recently demonstrated using a diamond memory via a Raman transition, and with a conversion efficiency of only 1\%~\cite{Fisher2016}. 
The system proposed here can finally allow the storage of photons as short as 100\,ps and convert them to much longer single photon pulses up to $\tau=1/\gamma_{-}\simeq100\,\mathrm{ns}$. 


\section{Conclusion}

We have proposed a new generation of devices for single photon production and manipulation by coupling two quantum emitters in a cavity. This is achieved by a dynamical control of the subradiant state which, when the emitters are detuned, acts as an effective dipole that can be externally modified. 

These concepts are ready to be implemented thanks to the last decades of experimental effort in the quantum-dot cavity systems, that have allowed to obtain near-optimal single photon sources in the solid state with impressive values regarding efficiency, brightness and indistinguishably~\cite{Ding2016,Somaschi2016,Senellart2017}.
These technological efforts can now lead to the on demand, efficient, single photon generation, with a tunable bandwidth on the range of $\left[\mathrm{10MHz,10GHz}\right]$. They can be used to expand the range of application of quantum light both in quantum computing and quantum sensing, with efficient single photon sources with continuous variable encoding in the time-frequency domain. The same architecture can also store broadband single photons, with pulse durations as low as 100\,ps, stored during 100\,ns, with an efficiency up to 68\%. Both storage and bandwidth control would finally allow efficient bandwidth conversion of single photons. 

Overall, the proposed device acts as a junction useful in any kind of hybrid quantum network, with the possibility to emit, store, and modify single photons of a wide variety of quantum systems. They also open new possibilities for both quantum sensing~\cite{andrewhypercube} and quantum computing~\cite{timefrequencyCVcomputing}, with inherently high efficiencies, as required for scalability.





\begin{acknowledgments}
The authors thank Andrew White, Perola Milman, Carlos Anton, Philippe Grangier and Riccardo Messina for fruitful discussions. This work was supported by SAFRAN-IOGS chair on Ultimate Photonics and a public grant overseen by the French National Research Agency (ANR) as part of the Investissements d'Avenir programme (Labex NanoSaclay, reference: ANR-10-LABX-0035). J.-J.G. acknowledges the support of Institut Universitaire de France. 
\end{acknowledgments}


%

\end{document}